\documentclass[twocolumn,showpacs,preprintnumbers,amsmath,amssymb]{revtex4}

\usepackage{epsfig}
\usepackage{graphics}
\usepackage{graphicx}
\usepackage{dcolumn}
\usepackage{bm}

\begin{document}

\preprint{APS/123-QED}

\title{Anticipating, Complete and Lag Synchronizations in RC Phase-Shift Network Based Coupled Chua's Circuits without Delay}

\author{K.~Srinivasan$^1$}
\author{D.~V.~Senthilkumar$^{2,3}$}
\author{I.~Raja Mohamed$^{4}$}
\author{K.~Murali$^{5}$}
\author{M.~Lakshmanan$^1$}%
 \author{J.~Kurths$^{3,6,7}$}
\affiliation{
$^1$Centre for Nonlinear Dynamics, Department of Physics, Bharathidasan University, Tiruchirapalli 620024, India\\
$^2$Centre for Dynamics of Complex Systems, University of Potsdam, 14469 Potsdam, Germany\\
$^3$Potsdam Institute for Climate Impact Research, 14473 Potsdam, Germany\\
$^4$Department of Physics, B.S.Abdur Rahman University, Chennai-600048, India\\
$^5$Department of Physics, Anna University, Chennai, India\\
$^6$Institute of Physics, Humboldt University, 12489 Berlin, Germany\\
$^7$Institute for Complex Systems and Mathematical Biology, University of Aberdeen, UK\\
}

\date{\today}

\begin{abstract} 

We construct a new RC phase shift network based Chua's circuit, which exhibits a period-doubling bifurcation route to chaos.  Using coupled versions of such a phase-shift network based Chua's oscillators, we describe a new method for achieving complete synchronization (CS), approximate lag synchronization (LS) and approximate anticipating synchronization (AS) without delay or parameter mismatch.  Employing the Pecora and Carroll approach, chaos synchronization is achieved in coupled chaotic oscillators, where the drive system variables control the response system.  As a result, AS or LS or CS is demonstrated without using a variable delay line both experimentally and numerically.

\end{abstract}

\maketitle

{\bfseries Synchronization of coupled chaotic systems is a fundamental nonlinear phenomenon observed in diverse areas of science and technology. Since its detection, different kinds of
synchronizations have been demonstrated both theoretically and experimentally.
The existence and/or transition between different kinds of synchronization
in a single coupled system  have also
been reported by tuning certain system parameters.  In particular, transitions
between anticipatory, complete and lag synchronizations have been demonstrated
in dynamical systems described by both ordinary and delay differential equations
by tuning the delay coupling and also in systems with parameter mismatch
without delay coupling. In this investigation, we will demonstrate
the existence of all the above three types of synchronizations in coupled
RC phase-shift network based Chua's circuits by using the Pecora and Carroll method without any parameter mismatch or delay coupling both experimentally (by
using electronic circuits) and theoretically (by simulating the normalized evolution equations).  The novelty of our approach is that we introduce a RC phase-shift network circuit to the coupled Chua's circuit which results in complete, lag and anticipating synchronizations depending upon the drive variable.  The method is particularly simple and elegant to implement and control. Just by simply switching the connection of response circuit with the drive system variable, different kinds of synchronization is shown to result in.

}

\section{\label{sec:level1}INTRODUCTION}

Synchronization of chaotic oscillations has been an area of extensive research since the pioneering works of Fujisaka and Yamada~\cite{fujisaka83} and of Pecora and Carroll~\cite{pecora90}.  Chaos synchronization properties of uni- or bidirectionally coupled chaotic systems have attracted the attention of many researchers due to their potential applications in a variety of fields~\cite{lakmurbook,pikovsky01}.  Apart from identical or complete synchronization (CS), other important forms of synchronization have also been identified~\cite{pikovsky01,srini10,srini11}.  Among other forms of  interesting types are the lag~\cite{rosenblum97} and anticipating synchronizations~\cite{voss00,voss01,pyragas08}, where coupled systems follow identical phase space trajectories but shifted in time relative to each other.  The anticipating and lag synchronization have been observed in lasers~\cite{masoller01,wu03}, neuronal models~\cite{li04,ciszak04}, and electronic circuits~\cite{taherion99,voss02,pethel03,blakely08,srini11}.  

By using an explicit time delay or memory both lag and anticipating synchronizations between unidirectionally coupled oscillators can be obtained~\cite{lakdvsbook}.  While lag synchronization is acheived by coupling the response system to a past state of the drive, anticipating synchronization can be obtained by a feedback control using the current state of the drive compared to the past state of the response~\cite{voss00,voss01}.  In both cases, an explicit time delay appears in the coupling.  In particular, the transition between anticipatory, complete and lag synchronizations has been demonstrated in dynamical systems described by delay differential equations by tuning the delay coupling~\cite{dvs05,srini11}. 
Another way to achieve approximate lag synchronization in mutually coupled chaotic oscillator is by using parameter mismatch~\cite{rosenblum97}.  Notably, intermittent and continuous lag synchronizations have been observed as intermittent steps in a route from phase to complete synchronization by increasing the coupling strength~\cite{rosenblum97,taherion99}.  In general, both lag and anticipating synchronizations with some finite amplitude error in unidirectionally coupled chaotic oscillators can be achieved using specific intentional parameter mismatch between the drive and the response systems~\cite{corron05}.  Also a new method for estimating the correlation and time shift between drive and response oscillators, using a new coupling scheme and linear filter theory, has been demonstrated in Ref.~\cite{blakely08}.  Hence, it is of interest to investigate other potential simple methods which can identify lag/anticipating synchronizations similar to the above procedures (that is time delay or parameter mismatch).

In this connection, Chua's circuit and its variants are well known chaotic circuits that exhibit a wide variety of nonlinear dynamics phenomena, such as bifurcations and chaos~\cite{chuabook,madan,lakmurbook,mlsrbook,kenn1992,chukom,chenueta,ogo}.  Pecora and Carroll have proposed that a subsystem of a chaotic system can be synchronized with a separate chaotic system under certain conditions~\cite{pecora90,pecora91}.  This idea of synchronization has been successfully applied to a variety of nonlinear systems  including phase-locked loops, hysteresis circuits etc. \cite{pecora90,pecora91,desousa,endo}.  In this paper, we have constructed a well-known simple RC phase shift network~\cite{hosokawa01} based Chua's circuit, which exhibits a period-doubling bifurcation route to chaotic attractor.  Then, we present a new method for achieving complete, lag and anticipating synchronizations in unidirectionally coupled chaotic Chua's oscillators.  In this method, by switching the parameter in the drive system, the response system is shown to exhibit CS or LS or AS, where neither a time delay nor a parameter mismatch is necessary.  In short, the mechanism proposed in this paper is a very simple and elegant way of achieving different types of synchronizations in unidirectionally coupled chaotic oscillators.  This method can also be extended to various applications including signal processing, temporal pattern recognition, secure communication and cryptography.  We demonstrate complete, lag and anticipating synchronizations in the designed circuit both numerically and experimentally.  

The organization of this paper is as follows.  In Sec. 2, the circuit realization of the RC phase shift network based Chua's circuit is presented, while in Sec. 3, the dynamics of the modified Chua's circuit is given. In Sec. 4, we discuss the three types of synchronizations (complete, lag and anticipating) exhibited by a single set of unidirectionally coupled Chua's circuits of the above type through appropriate switching. The paper concludes with a summary in Sec. 5.

\section{\label{sec:level2}CIRCUIT REALIZATION}

\subsection{Circuit Design}
The standard Chua's circuit~\cite{madan,lakmurbook} contains an LC oscillator connected to a nonlinear element, namely a Chua's diode through an RC circuit.  The possibility of constructing an n-dimensional Chua's circuit using either an LC or an RC ladder network is indicated in Ref.~\cite{gotz}.
In this paper we design and implement a phase shift network of modified Chua's circuit by introducing three RC circuits in it as in Fig.~\ref{chua_m_cir}.  Each RC circuit introduces a finite phase shift with an attenuation of the signal.  The main objective to design such a circuit is to study different types of chaotic synchronizations in a simple and elegant way without the introduction of an explicit time delay or parameter mismatch.  Two or more circuits of this kind can be coupled to form a network and this coupling is made by connecting any one of the three RC circuits of the drive system to the response system.  We find that depending upon the choice of the RC circuit used for coupling, the system exhibits different kinds of synchronization and this is the prime reason for introducing the RC circuits for phase shift.  

\subsection{Circuit Equations}
\begin{figure} 
\centering
\includegraphics[width=1.0\columnwidth]{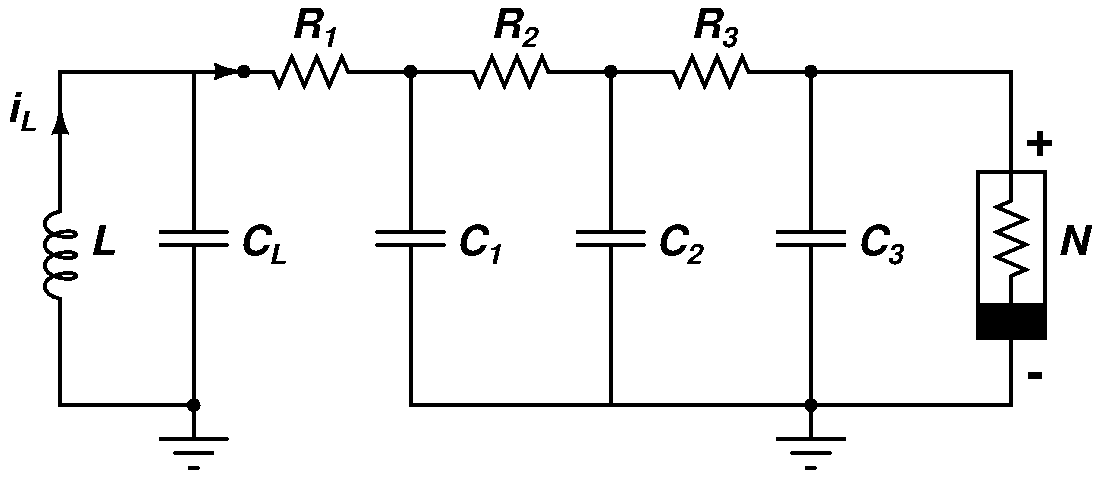}
\caption{\label{chua_m_cir} Circuit realization of phase shift network based Chua's circuit.  Here, $N$ is Chua's diode.  The
parameter values of the other elements are fixed as $L=18.0~mH$,
$C_1 = C_2 = C_3=4.7~nF$, $C_L=100.0~nF$ and $R_1 = R_2 = R_3 = 610~\Omega$.}
\end{figure}

The RC phase shift network based Chua's circuit is shown in Fig.~\ref{chua_m_cir}.  It contains
four capacitors $C_1, C_2, C_3$ and $C_L$, an inductor $L$, three linear resistors $R_1, R_2$ and $R_3$ and only one nonlinear element, namely Chua's diode $(N)$.

By applying Kirchhoff's laws to this circuit
\cite{lakmurbook}, the governing equations for the voltage $v_1$ across the
capacitor $C_1$, voltage $v_2$ across the capacitor $C_2$, voltage $v_3$ across the capacitor $C_3$, voltage $v_{LC}$ across the capacitor $C_L$ and the current $i_L$ through the inductor $L$ are given by the following set of five coupled first-order autonomous nonlinear (piecewise) differential equations
\begin{subequations}
\begin{eqnarray} 
C_1\frac{dv_1}{dt} &=& \frac{1}{R_1}(v_{LC}-v_1) + \frac{1}{R_2}(v_2-v_1), \\
C_2\frac{dv_2}{dt} &=& \frac{1}{R_2}(v_1-v_2) + \frac{1}{R_3}(v_3-v_2), \\
C_3\frac{dv_3}{dt} &=& \frac{1}{R_3}(v_2-v_3) - i_N, \\
C_L\frac{dv_{CL}}{dt} &=& \frac{1}{R_1}(v_1-v_{LC}) + i_L,  \\
L\frac{di_L}{dt} &=&-v_{LC}.
\end{eqnarray} 
\label{cir_eqn}
\end{subequations}
The term $i_N=f(v_3)$ represents the $v - i$
characteristics of Chua's diode and is
given by
\begin{eqnarray}
f(v_3) &=& G_b v_3+0.5(G_a-G_b) [|v_3+B_p|-|v_3-B_p|], 
\end{eqnarray}
where $G_a$ and $G_b$ are the inner and outer slopes of the characteristic curve
respectively.   Here $\pm B_p$ denotes the break point of the characteristic
curve.  The experimental parameters of the circuit  elements are fixed at $L=18.0~mH$, $C_1=C_2=C_3=4.7~nF$, $C_L=100.0~nF$, $R_1=R_2=R_3=610~\Omega$, 
$G_a=-0.76~mS, G_b=-0.41~mS$ and $\pm B_p=\pm1.0~V$.

Eq.~(\ref{cir_eqn}) can be converted into a normalized form, convenient for numerical analysis by using
the following rescaled variables and parameters, $v_1=x B_p$, $v_2=y B_p$, $v_3=z B_p$, $v_{LC}=\omega B_p$,
$i_L=(B_p G)h$, $t=C_L t' /G$, $G=1/R_3$.  Note that here $t'$ is in dimensionless unit.  The set of normalized equations so obtained are
\begin{subequations}
\begin{eqnarray} 
\dot{x} &=& \beta_1(\omega-x)+\beta_2(y-x), ~~~\left(\dot{} = \frac{d}{dt'}\right) \\
\dot{y} &=& \beta_3(x-y)+\beta_4(z-y), \\
\dot{z} &=& \alpha[(y-z)-g(z)], \\
\dot{\omega} &=& \gamma (x-\omega)+h, \\          
\dot{h} &=& -\beta_0 \omega,
\end{eqnarray}
\label{nor_eqn}
\end{subequations}
where $\alpha = (C_L/C_3R_3G)$, $\gamma=1/(GR_1)$, $\beta_0 = C_L/(LG^2)$, $\beta_1 = C_L/(C_1GR_1)$, $\beta_2 = C_L/(C_1GR_2)$, $\beta_3 = C_L/(C_2GR_2)$ and $\beta_4 = C_L/(C_2GR_3)$.  The term $g(z)$ is obviously represented in
the rescaled form as 
\begin{equation}
g(z) = bz + 0.5(a - b) [|z + 1| - |z - 1|].  
\end{equation}
\\
Here, $a=G_a/G$ and $b=G_b/G$.  The dynamics of Eq.~(\ref{nor_eqn}) now depends on
the rescaled parameters $\alpha$, $\gamma$, $\beta_0$, $\beta_1$, $\beta_2$, $\beta_3$, $\beta_4$, $a$ and $b$.  The parameter values
are fixed as $\alpha=\beta_1=\beta_2=\beta_3=\beta_4=21.2765$, $\gamma=1$, $a=-0.462$ and $b=-0.25$, while varying $\beta_0$.  

\begin{figure} 
\centering
\includegraphics[width=0.8\columnwidth]{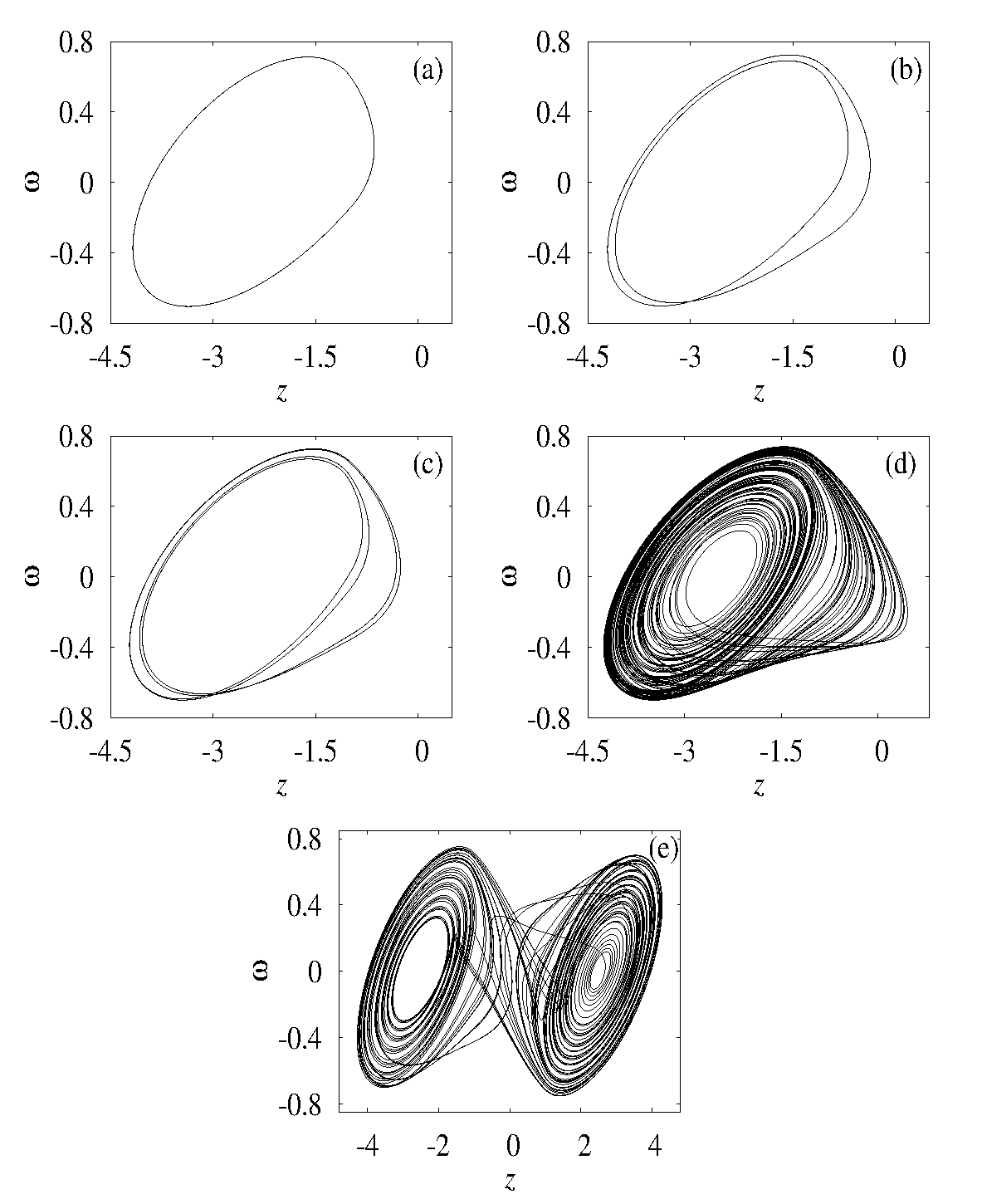}
\caption{\label{chm_pha_num} Phase portraits of Eq.~(\ref{nor_eqn}). (a) $\beta_0=3.383$ (Period-T), (b) $\beta_0=3.2356$ (Period-2T), (c) $\beta_0=3.194$ (Period-4T), (d) $\beta_0=3.05$ (One-band chaos) and (e) $\beta_0=2.953$ (Double-band chaos).  } 
\end{figure}  
\section{\label{sec:level41}DYNAMICS OF RC PHASE SHIFT NETWORK BASED CHUA'S CIRCUIT}

\begin{figure} 
\centering
\includegraphics[width=0.7\columnwidth]{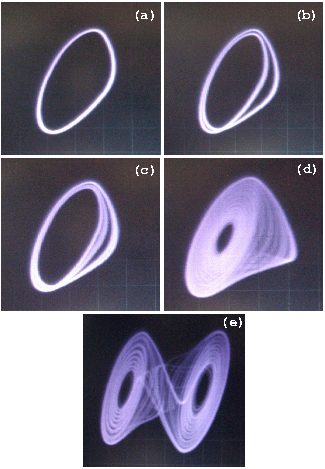}
\caption{\label{chm_pha_exp} The corresponding experimental results of Figs.~\ref{chm_pha_num}. Vertical scale $2v/div$., horizontal scale $0.5v/div$.} 
\end{figure}

In this section, we first present the results of our numerical study of system (\ref{nor_eqn}) so as to make the underlying dynamics clear and then present the corresponding experimental results of the associated circuit (Fig.~\ref{chua_m_cir}) described by (\ref{cir_eqn}).  We fix all the parameter values as mentioned in the previous section.  From the nature of the numerical results obtained by solving Eq.~(\ref{nor_eqn}), using the
standard fourth order Runge-Kutta algorithm, we infer the following picture.  We use the system parameter $\beta_0$ as the control parameter.  When $\beta_0$ is varied from $3.381$ downwards, the system exhibits the familiar period-doubling bifurcation route to chaos, followed by periodic windows, etc.  In addition, a few other interesting dynamical phenomena are also identified by a careful study through $\beta_0$ scanning.  This is illustrated in 
Figs.~\ref{chm_pha_num} in the ($z - \omega$) phase plane.  Experimentally, the phase trajectory is obtained by measuring the voltage levels $v_3$ and $v_{LC}$ in the circuit of Fig.~\ref{chua_m_cir} and connected to the $X$ and $Y$ channels of an oscilloscope.  The phase trajectory so obtained is shown in Figs.~\ref{chm_pha_exp}.  Similar to numerical studies, experiments reveal a transition from periodic attractor to chaotic attractor through universal period doubling route.   

\begin{figure} 
\centering
\includegraphics[width=1.0\columnwidth]{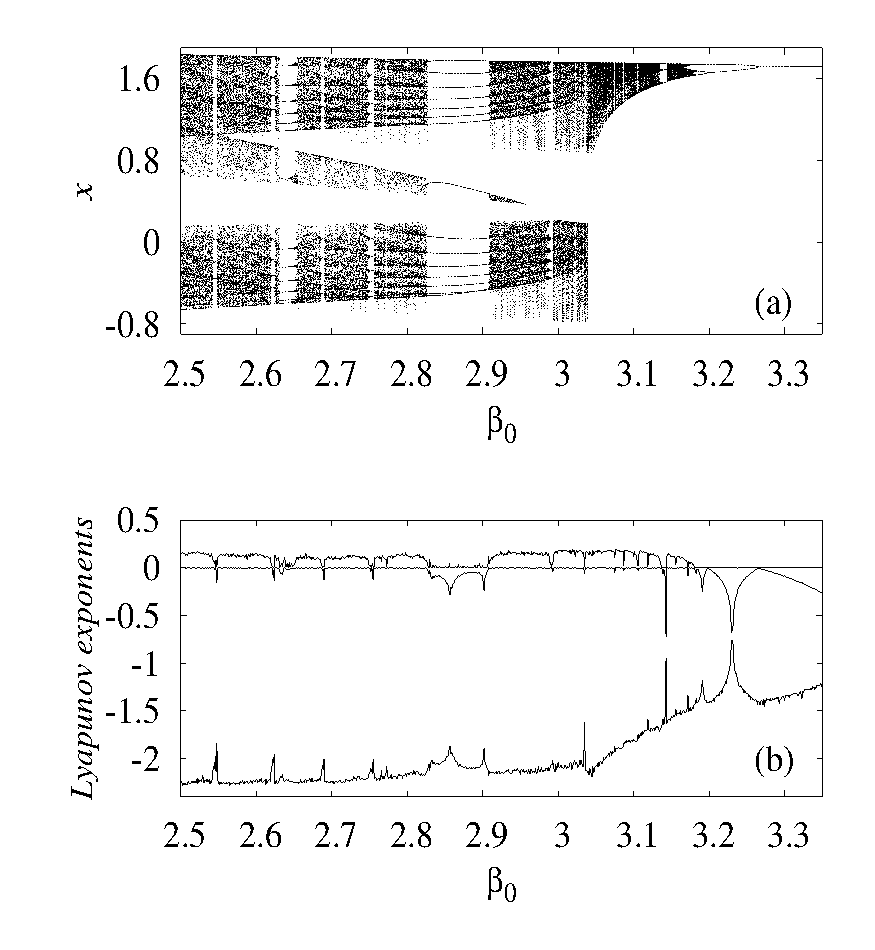}
\caption{\label{chum_bif_lyp} One parameter bifurcation diagram $(\beta_0 - x)$ for the parameter values $\alpha=\beta_1=\beta_2=\beta_3=\beta_4=21.2765$, $\gamma=1$, $a=-0.462$ and $b=-0.25$.} 
\end{figure}

The above details can be easily inferred from the one parameter bifurcation diagram in the ($\beta_0 - x$) plane and the corresponding three maximal Lyapunov exponents in the ($\beta_0 - \lambda_m$), $m=1,2,3$, plane associated with Eq.~(\ref{nor_eqn}) which are given in Figs.~\ref{chum_bif_lyp}.  In particular, the standard period-doubling bifurcation sequence to chaos and windows have been observed for a range of parameter values, $3.0 <\beta_0 < 3.5$.  For example, it is clear that for $\beta_0 > 3.3$ there is a limit cycle attractor of period-$T$.  At $\beta_0 = 3.26$, a period doubling bifurcation occurs and a period-$2T$ limit cycle develops and is stable in the range $3.26>\beta_0>3.2$.  When $\beta_0$ is decreased further the period-$2T$ limit cycle bifurcates to a period-$4T$ ($3.19>\beta_0>3.18$) attractor.  Further period doubling occurs when $\beta_0<3.18$ giving rise to $8T$ and $16T$ period limit cycle, respectively.  The chaotic attractor (single band) is first observed at $\beta_0 = 3.175$.  Further decrease in the parameter ($\beta_0$) of the system causes it to admit double band chaotic nature.  For $3.0>\beta_o>2.5$, the dynamics is even more complicated and intricate.  This interval of $\beta_0$ is not fully occupied by chaotic orbits alone. Many fascinating changes in the dynamics like reverse period-doubling bifurcations, periodic orbits (windows), period-doubling of windows, intermittency and antimonotonicity   
take place at different critical values of $\beta_0$ in this range.
\begin{figure} 
\centering
\includegraphics[width=1.0\columnwidth]{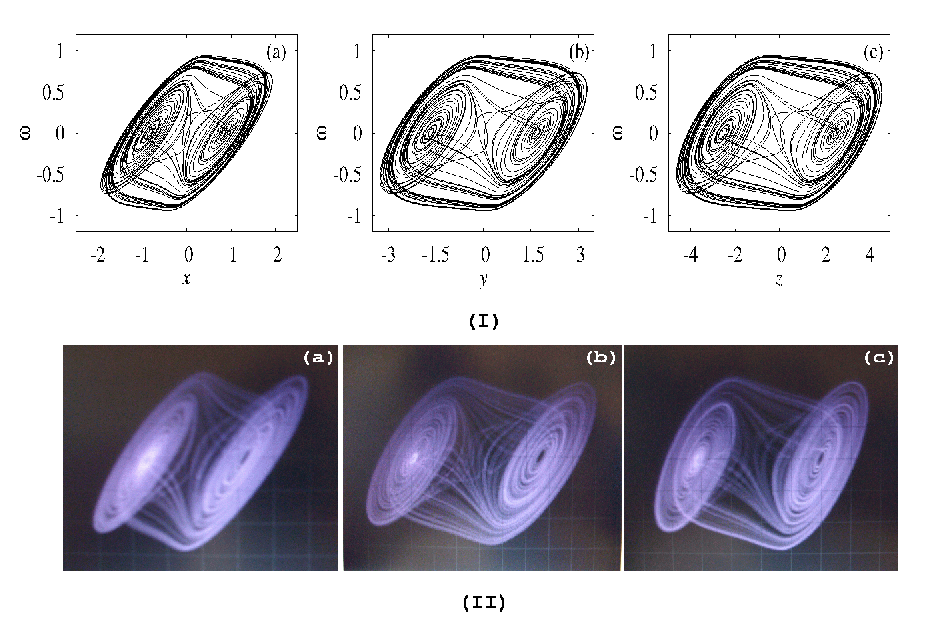}
\caption{\label{chum_pha_2b1} (I) Double band chaotic attractors of the system (\ref{nor_eqn}): (a) ($x- \omega$) plane, (b) ($y - \omega$) plane and (c) ($z -\omega$) plane.  (II) Phase portraits (experimental) of chaotic attractor from the circuit (Fig.~\ref{chua_m_cir}), (a) ($v_1-v_{LC}$); vertical scale $1v/div.$, horizontal scale $0.5v/div$., (b) ($v_2-v_{LC}$); vertical scale $1v/div.$, horizontal scale $0.5v/div$. and (c) ($v_3-v_{LC}$); vertical scale $2v/div.$, horizontal scale $0.5v/div$.}
\end{figure}  

\begin{figure} 
\centering
\includegraphics[width=1.0\columnwidth]{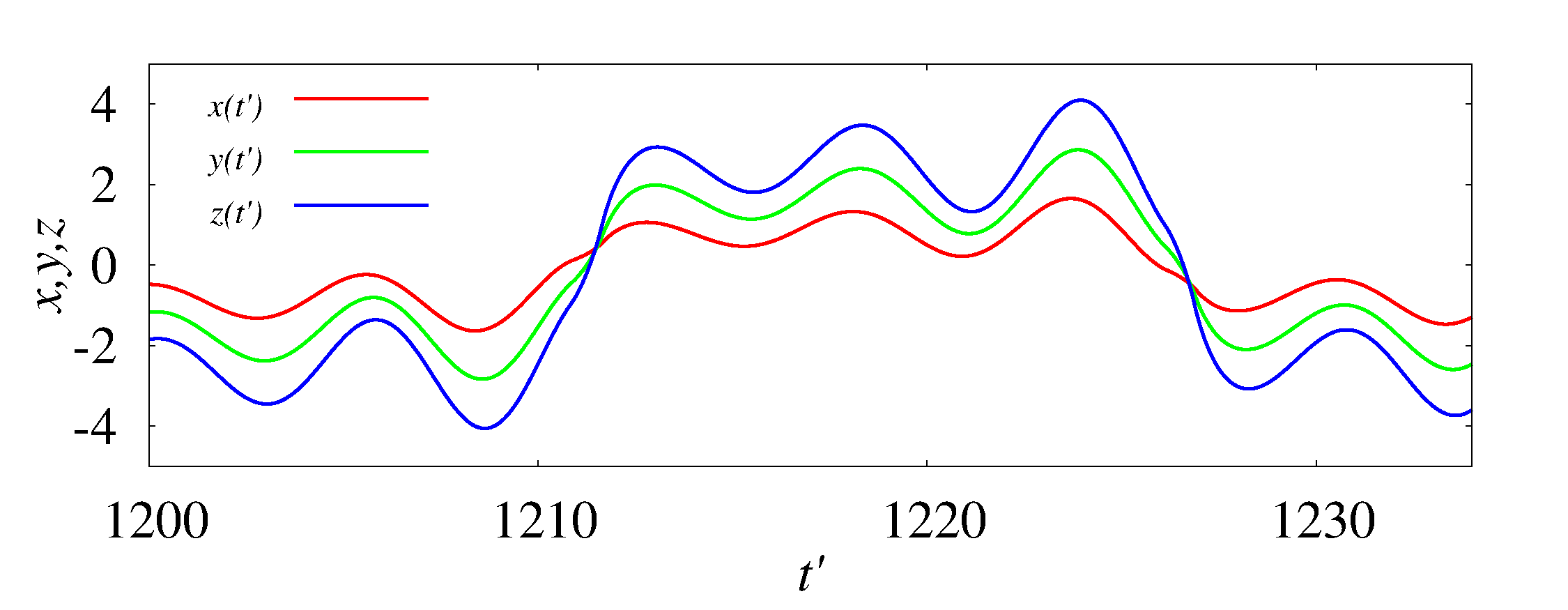}
\caption{\label{chum_ph_time} Time-series plot for double band chaotic attractor of system (\ref{nor_eqn}). Note the phase shift between the three dynamical variables $x$, $y$ and $z$.} 
\end{figure}

At $\beta = 2.067$, the system (\ref{nor_eqn}) exhibits double band chaotic attractor which is shown in Fig.~\ref{chum_pha_2b1}(I) for different projections of phase space and the corresponding experimental analysis is shown in Fig.~\ref{chum_pha_2b1}(II).  For the same $\beta$ value, the time series plot is presented in Fig.~\ref{chum_ph_time}.  The phase shift in $x$, $y$ and $z$ can be clearly seen in Fig.~\ref{chum_ph_time} due to the three RC phase shift circuits shown in Fig.~\ref{chua_m_cir}. 

\section{\label{sec:level3} COUPLED RC PHASE SHIFT NETWORK BASED CHUA'S CIRCUITS: LAG AND ANTICIPATING SYNCHRONIZATIONS}

\begin{figure}
\centering
\includegraphics[width=1.0\columnwidth]{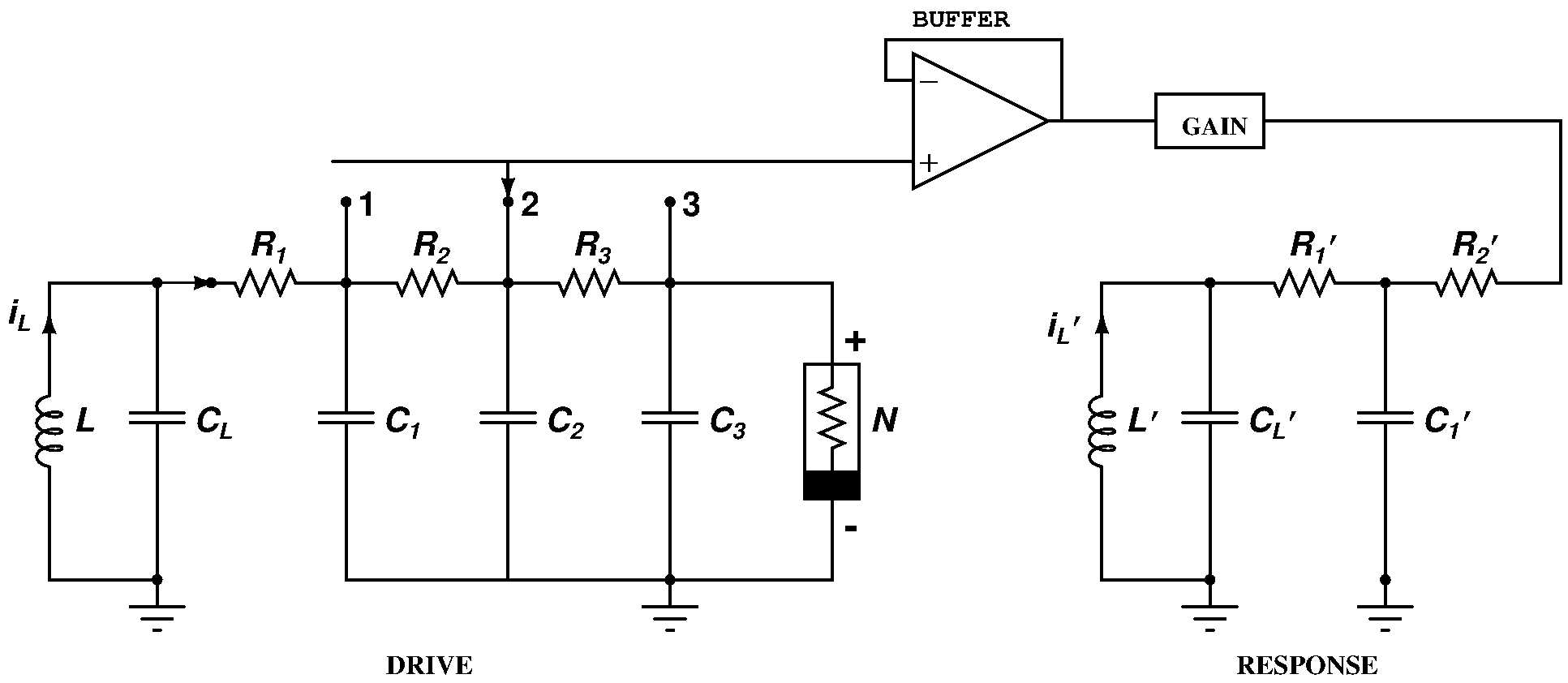}
\caption{\label{chuam_coup_cir} Circuit realization of coupled RC phase shift network based Chua's circuit.}
\end{figure}

Next, we study the dynamics of coupled phase-shift network based Chua's circuits which is shown in Fig.~\ref{chuam_coup_cir}.  The network is made by connecting the drive circuit to the response circuit through a buffer and a gain amplifier.  Here, the single phase-shift network based Chua's circuit acts as the drive circuit and the governing circuit equation for the drive part is nothing but Eq.~(\ref{cir_eqn}).  Depending upon the value of the feedback resistance of the standard inverting amplifier, the gain can be fixed.  Using the Pecora and Carroll approach of building an identical copy of the response subsystem, we demonstrate three types of chaos synchronization, namely complete, lag and anticipating synchronizations in the proposed Chua's circuit both experimentally and numerically without the introduction of any time-delay or parameter mismatch.  When the voltage across $C_2$ is used to drive the subsystem, complete synchronization is observed. On the other hand when voltage across $C_1$ and $C_3$ are used, lag and anticipating synchronizations are observed. By simply connecting (switching) the response system to the drive system through either of the three terminals $1$, $2$ and $3$ (shown in Fig.~\ref{chuam_coup_cir}) we observed three types of synchronizations.  The reason for introducing the different coupling of state variables $v_1$, $v_2$ and $v_3$ is to observe all the three types of synchronization, namely lag, identical and anticipating synchronizations, in a simple manner.  This is achieved by exploiting the finite phase-shift that is being introduced by the individual RC network element.

\subsection{Dynamical systems}
The governing circuit equations are given below.\\
(a) Drive system : \\
Same as Eq.~(\ref{cir_eqn}), including the circuit parameter values. \\
(b) Response system : \\
\begin{subequations}
\begin{eqnarray} 
C_1'\frac{dv_1'}{dt} &=& \frac{1}{R_1'}(v_{LC}'-v_1') + \frac{1}{R_2'}(\epsilon \times v_2-v_1'), \\
C_L'\frac{dv_{CL}'}{dt} &=& \frac{1}{R_1'}(v_1'-v_{LC}') + i_L',  \\
L'\frac{di_L'}{dt} &=&-v_{LC}'.
\end{eqnarray}
\label{cop_cir_eqn}
\end{subequations} 
The term $i_N=f(v_3)$ represents as before the $v - i$ characteristics of the  Chua's diode.  The parameters of the circuit  elements are fixed at  $L'=18.0~mH$, $C_1'=4.7~nF$, $C_L'=100.0~nF$, $R_1'=610~\Omega$ and $R_2'=610~\Omega$.  Here $\epsilon$ is the gain term.  Eqs.~(\ref{cop_cir_eqn}) are now rescaled as follows: with $v_1'=x' B_p$, $v_{LC}'=\omega' B_p$, $i_L'=(B_p G)h'$, $t=C_L t' /G$.  Then the rescaled version of the equation is given as follows \\
(a) Drive system : \\
Same as Eq.~(\ref{nor_eqn}), including the parameter values.\\
(b) Response system : \\
\begin{subequations}
\begin{eqnarray} 
\dot{x'} &=& \beta_1'(\omega'-x')+\beta_2'(\epsilon' \times y-x'), \\
\dot{\omega'} &=& \gamma' (x'-\omega')+h', \\
\dot{h'} &=& -\beta_0' \omega',
\end{eqnarray}
\label{cop_nor_eqn}
\end{subequations}
where $\beta_0' = C_L'/(L'G^2)$, $\beta_1' = C_L'/(C_1'GR_1')$, $\beta_2' = C_L'/(C_1'GR_2')$, $\gamma'=1/(GR_1')$ and $\epsilon'=\epsilon$.  The parameter values become (due to the above choice of the circuit parameters) $\beta_1'=\beta_2'=21.2765$, $\gamma'=1.0$ and $\beta_0'=2.0672$.

\begin{figure}
\centering
\includegraphics[width=1.0\columnwidth]{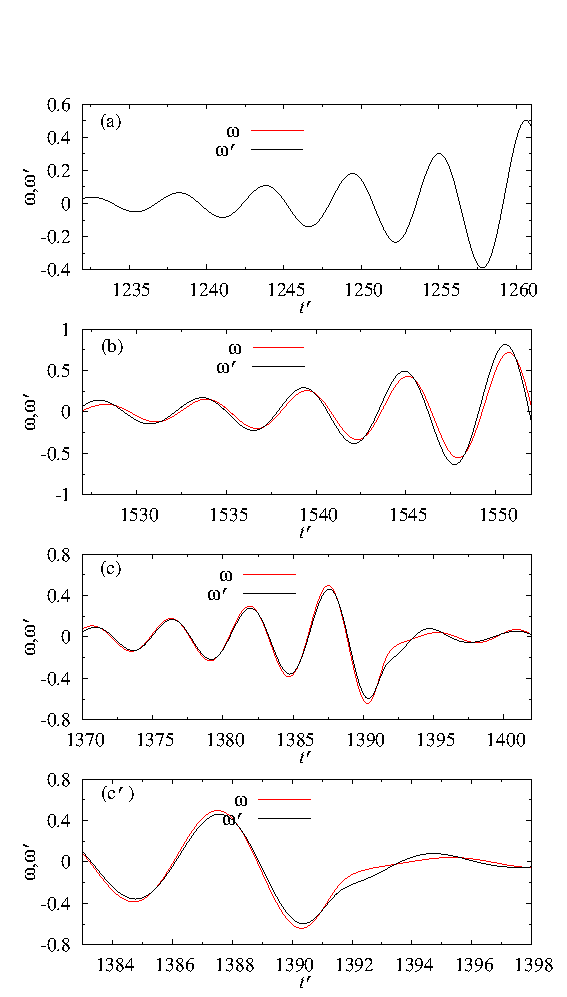}
\caption{\label{num_syn3_time}Numerically observed drive (red, $\omega$) and response (black, $\omega'$) wave forms showing: (a) identical synchronization, (b) approximate anticipating synchronization, (c) approximate lag synchronization in coupled chaotic oscillators and (c') enlarged version of Fig.~\ref{num_syn3_time}(c).} 
\end{figure}  

\subsection{Chaos synchronization}
\begin{figure}
\centering
\includegraphics[width=0.6\columnwidth]{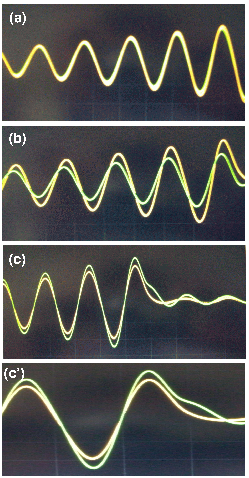}
\caption{\label{exp_syn3_time} Experimentally observed drive (yellow, $v_{LC}$) and response (green, $v_{LC}'$) wave forms showing: (a) identical synchronization, (b) approximate anticipating synchronization, (c) approximate lag synchronization in coupled chaotic oscillators and (c') enlarged version of Fig.~\ref{exp_syn3_time}(c).  Vertical scale $0.5v/div.$: horizontal scale $200\mu s/div$.} 
\end{figure}

\begin{figure}
\centering
\includegraphics[width=0.6\columnwidth]{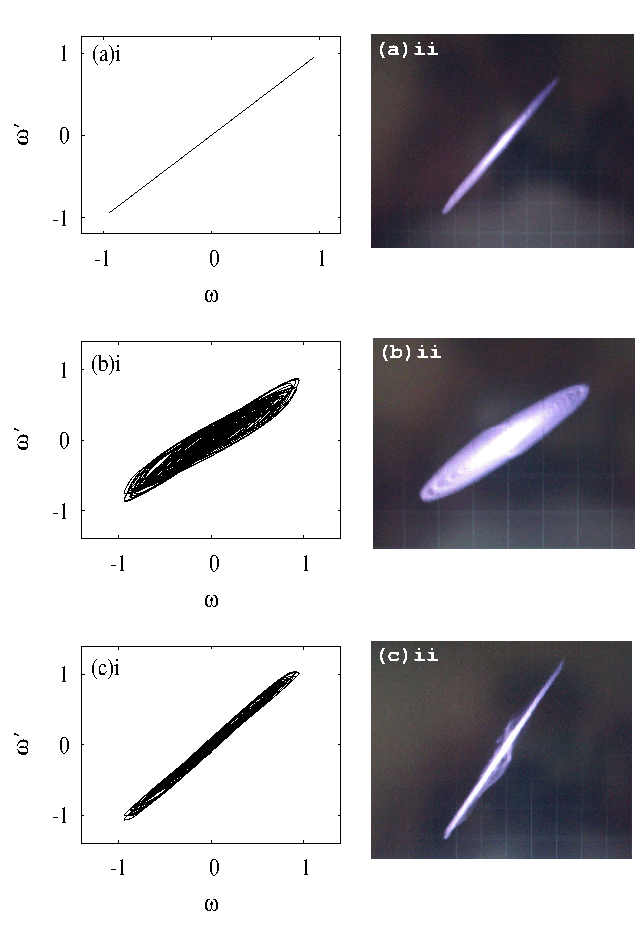}
\caption{\label{chum_syn3_pha} (i) Numerically observed drive ($\omega$) and response ($\omega'$) phase plane in $(\omega-\omega')$: (a) identical synchronization, (b) approximate anticipating synchronization and (c) approximate lag synchronization in coupled chaotic oscillators.  (ii) Experimentally observed ($v_{LC}$) and response ($v_{LC}'$) phase plane in ($v_{LC}$-$v_{LC}'$).  Vertical scale $1v/div$., horizontal scale $1v/div$.}  
\end{figure}

The unidirectionally coupled RC phase shift network based Chua's circuit (Fig.~\ref{chuam_coup_cir}) shows that the response subsystem contains an identical oscillator as that of the drive system.  Here, the drive system controls the response, through the drive component in Eq.~(\ref{cop_cir_eqn}a) and Eq.~(\ref{cop_nor_eqn}a).  Connecting the response system to the terminal $2$ (see Fig.~\ref{chuam_coup_cir}) of the drive system, we observed complete synchronization.  With $\epsilon=1.0$ and the $y$-drive component coupled through one way coupling with the response subsystem, the coupled oscillators exhibit identical synchronization.  Time series of the drive and response variables ($\omega$ \& $\omega'$) are shown in Fig.~\ref{num_syn3_time}$(a)$ and the corresponding experimental plot is shown in Fig.~\ref{exp_syn3_time}$(a)$.  In Fig.~~\ref{exp_syn3_time} the horizontal axis is calibrated as $200~\mu s/div.$ and the vertical axis is $0.5~v/div$.  From these figures, it is clearly seen that both the phase and amplitude of the drive and response systems are coinciding, proving that the coupled system of Fig.~\ref{chuam_coup_cir} exhibits complete synchronization.  The same is observed in the phase space plots, shown in Fig.~\ref{chum_syn3_pha}$(a)$.  

\begin{figure}
\centering
\includegraphics[width=0.8\columnwidth]{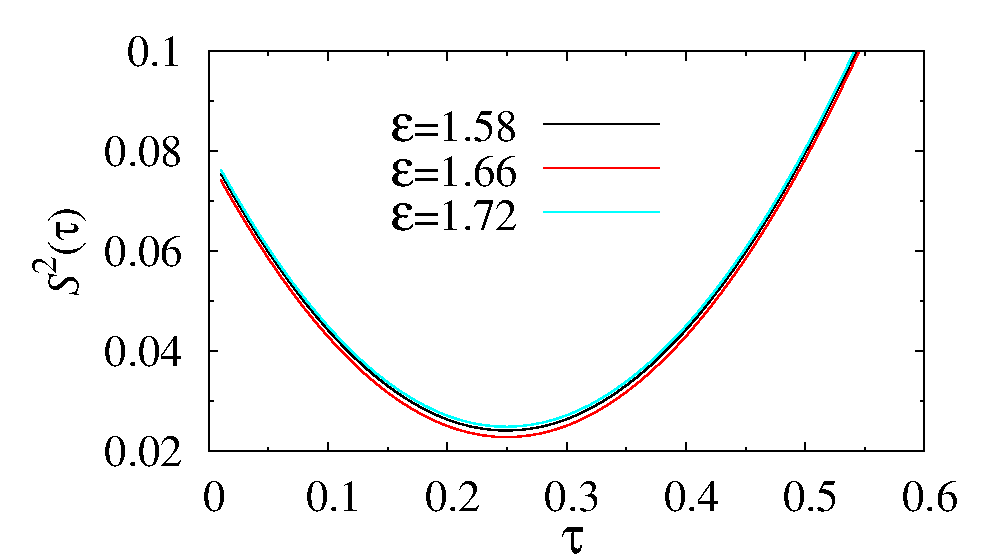}
\caption{\label{anti_sim} Similarity function $S^2(\tau)$  corresponding to Fig.~\ref{num_syn3_time}(b) confirming
anticipatory synchronization for different values of $\epsilon$ (red, $\epsilon=1.66$; black, $\epsilon=1.58$; cyan, $\epsilon=1.72$).}  
\end{figure}

Now, connecting the response system to the terminal $1$ (see Fig.~\ref{chuam_coup_cir}) of the drive system variable $v_1$ is coupled with the response subsystem and correspondingly Eq.~(\ref{cop_cir_eqn}a) and Eq.~(\ref{cop_nor_eqn}a) get modified respectively as
\begin{equation} 
C_1'\frac{dv_1'}{dt} = \frac{1}{R_1'}(v_{LC}'-v_1') + \frac{1}{R_2'}(\epsilon \times v_1-v_1')
\end{equation}
and
\begin{equation} 
\dot{x'} = \beta_1'(\omega'-x')+\beta_2'(\epsilon' \times x-x').
\end{equation}
For $\epsilon=1.66$ and $x$-drive component coupled through one way coupling with the response subsystem, the coupled oscillators exhibit anticipatory synchronization.  Figures~\ref{num_syn3_time}$(b)$ and \ref{exp_syn3_time}$(b)$ depict the time series plot of $\omega$ and $\omega'$ numerically and experimentally.  From these figures we can observe that $\omega'$ anticipates $\omega$.  In other words, the response system anticipates the drive system, thereby we can infer the anticipatory synchronization of the system shown in Fig.~\ref{chuam_coup_cir}.  Anticipating synchronization can also be inferred from the phase space plot shown in Fig.~\ref{chum_syn3_pha}$(b)$.  The degree of synchronization with the corresponding time shift $\tau$ can be quantified using the similarity function~\cite{rosenblum97} defined as
\begin{equation}
S^2(\tau)=\frac{\langle [\omega'(t'-\tau)-\omega(t')]^2\rangle}{[\langle \omega^2(t')\rangle \langle \omega'^{2}(t') \rangle ]^{1/2}},
\label{anti_simi_eq}
\end{equation}
where $\langle \omega \rangle$ means the time average over the variable $\omega$.  If the signals $\omega (t')$ and $\omega^{'}(t')$ are independent, the difference between them is of the same order as the signals themselves.  If $\omega (t') = \omega^{'}(t')$, as in the case of complete synchronization, the similarity function reaches a minimum $S^2(\tau)=0$ for $\tau=0$.  But for the case of a nonzero value of time shift $\tau$, if $S^2(\tau)=0$, then there exists a time shift $\tau$ between the two signals $\omega (t')$ and $\omega^{'}(t')$ such that $\omega^{'}(t'-\tau)=\omega(t')$, demonstrating anticipating synchronization.  Figure \ref{anti_sim} shows the similarity function $S^2(\tau)$ as a function of the coupling delay $\tau$ for the three different values of $\epsilon$.  One may note that the minimum of $S^2(\tau)\approx 0.024$ occurs at $\tau \approx 0.25$ for $\epsilon=1.66$.  This indicates that there exists a time shift, corresponding to an anticipating time $\tau \approx 0.25$ time units, between the two signals in Fig.~\ref{num_syn3_time}(b) such that $\omega'(t'-\tau) \approx \omega(t')$ demonstrating approximate anticipactory synchronization. Translated into experimental units (see Sec. III) this time shift works out to be approximately $15.25\mu s$. The nearness of $S^2(\tau)$ to the value zero quantifies the degree of synchronization and hence $S^2(\tau)\approx 0.024$  attributes to the approximate anticipatory synchronization as observed in Figs.~\ref{num_syn3_time}(b) and \ref{exp_syn3_time}(b). It is also to be noted that for  slightly higher and lower values of $\epsilon$, the minimum of $S^2(\tau)$ occurs at the same $\tau$ but with further 
reduced degree of synchronization, indicated by their respective minima of $S^2(\tau)$, than that for $\epsilon=1.66$.

\begin{figure}
\centering
\includegraphics[width=0.8\columnwidth]{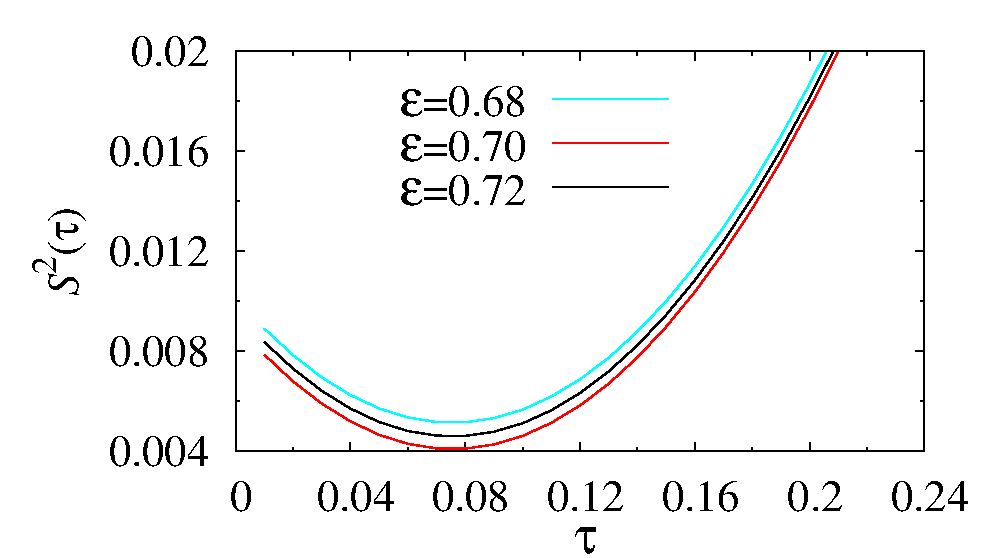}
\caption{\label{lag_sim}Similarity function $S^2(\tau)$ corresponding to Fig.~\ref{num_syn3_time}(c) confirming
lag synchronization for different values of $\epsilon$ (red, $\epsilon=0.7$; black, $\epsilon=0.72$; cyan, $\epsilon=0.68$).} 
\end{figure} 

Next, connecting the response subsystem to the terminal $3$ (see Fig.~\ref{chuam_coup_cir}) of the drive system variable $v_3$ is coupled and  correspondingly Eq.~(\ref{cop_cir_eqn}a) and Eq.~(\ref{cop_nor_eqn}a) get modified as
\begin{equation} 
C_1'\frac{dv_1'}{dt} = \frac{1}{R_1'}(v_{LC}'-v_1') + \frac{1}{R_2'}(\epsilon \times v_3-v_1')
\end{equation}
\begin{equation} 
\dot{x'} = \beta_1'(\omega'-x')+\beta_2'(\epsilon' \times z-x').
\end{equation}
For $\epsilon=0.7$ and the $z$-drive component coupled through one way coupling to the response subsystem, the coupled oscillators exhibit lag synchronization.  Numerical and experimental plots of time series of $\omega$ and $\omega'$ or $v_{LC}$ and $v_{LC}'$ are shown in Figs.~\ref{num_syn3_time}$(c)$ and \ref{exp_syn3_time}$(c)$.  In this case, the response system variable $\omega'$ lags the drive system variable $\omega$.  In other words the circuit (Fig.~\ref{chuam_coup_cir}) exhibits lag synchronization.  The phase space plots for lag synchronizations is shown in Fig.~\ref{chum_syn3_pha}$(c)$.  Further, in the present case also we can use the same similarity function $S^2(\tau)$ to characterize lag synchronization with a positive time shift $\tau$ instead of the negative time shift $\tau$ in Eq.~(\ref{anti_simi_eq}).  Figure~\ref{lag_sim} shows the similarity function $S^2(\tau)$ vs $\tau$ for three different values of $\epsilon$.  The minimum of similarity function becomes $S^2(\tau)\approx 0.004$ at $\tau \approx 0.078$ indicating that there is a time shift (Fig.~\ref{lag_sim}) between the drive and response signals $\omega (t')$ and $\omega^{'}(t')$, such that $\omega^{'}(t'+\tau)\approx\omega(t')$, confirming the occurrence of lag synchronization.  The minimum of $S^2(\tau)\approx 0.004$ corresponds to the approximate lag synchronization with lag time $\tau \approx 0.078$ between $\omega (t')$ and $\omega^{'}(t')$ in Fig.~\ref{num_syn3_time}$(c)$, which corresponds to an experimental lag time $4.76 \mu s$, approximately.

\section{\label{sec:level4}SUMMARY AND CONCLUSION}

In this paper,  we have constructed a three RC phase shift network based Chua's circuit, which exhibits a period-doubling bifurcation route to chaos.  Further, we have confirmed the generation of chaos by calculating the Lyapunov exponents and have investigated the related bifurcation phenomena.  Without introducing any time delay or parameter mismatch, different chaotic synchronizations are achieved by switching the drive system variables, using a one way coupling approach to the response subsystem.   We have explored the effectiveness of the approach using numerical simulations and the corresponding experimental results.  As a result, the complete, lag and anticipating synchronizations are controlled by the drive component on the response system.  It is worth mentioning that the three types of synchronization can also be realized for a similar identical response circuit with error feedback coupling, which will be discussed separately.  By extending the RC phase-shift networks cascadingly, one can realize a simple delayed chaotic circuit with the delay element simply made-up of RC networks.  Further, by exploiting phase synchronization in such networks, one can realize phase-synchronization based logic elements~\cite{murali07}. Such possibilities will be explored in future.

\begin{acknowledgments}
The authors are very grateful to an anonymous referee for some very valuable and critical comments which helped to improve the presentation of the results in Sec. IV.B.  The work of K.S. and M.L. has been supported by the Department of Science 
and Technology (DST), Government of India sponsored IRHPA research project, 
and DST Ramanna program and DAE Raja Ramanna program of M.L. 
D. V. S. and J. K. acknowledges the support from EU under Project
No. 240763 PHOCUS (FP7-ICT-2009-C).

\end{acknowledgments}

\newpage 


\begin{thebibliography}{49}
\expandafter\ifx\csname natexlab\endcsname\relax\def\natexlab#1{#1}\fi
\expandafter\ifx\csname bibnamefont\endcsname\relax
  \def\bibnamefont#1{#1}\fi
\expandafter\ifx\csname bibfnamefont\endcsname\relax
  \def\bibfnamefont#1{#1}\fi
\expandafter\ifx\csname citenamefont\endcsname\relax
  \def\citenamefont#1{#1}\fi
\expandafter\ifx\csname url\endcsname\relax
  \def\url#1{\texttt{#1}}\fi
\expandafter\ifx\csname urlprefix\endcsname\relax\def\urlprefix{URL }\fi
\providecommand{\bibinfo}[2]{#2}
\providecommand{\eprint}[2][]{\url{#2}}

\bibitem[{\citenamefont{Fujisaka and Yamada}(1983)\citenamefont{Fujisaka and Yamada}}]{fujisaka83}
\bibinfo{author}{\bibfnamefont{H.}~\bibnamefont{Fujisaka}}, \bibnamefont{and}
  \bibinfo{author}{\bibfnamefont{T.}~\bibnamefont{Yamada}}, 
  \bibinfo{journal}{Prog. Theor. Phys.} \textbf{\bibinfo{volume}{69}},
  \bibinfo{pages}{32} (\bibinfo{year}{1983}).

\bibitem[{\citenamefont{Pecora and Carroll}(1990)\citenamefont{Pecora and Carroll}}]{pecora90}
\bibinfo{author}{\bibfnamefont{L.~M} \bibnamefont{Pecora}}, \bibnamefont{and}
  \bibinfo{author}{\bibfnamefont{T.~L} \bibnamefont{Carroll}}, 
  \bibinfo{journal}{Phys. Rev. Lett.} \textbf{\bibinfo{volume}{64}},
  \bibinfo{pages}{821} (\bibinfo{year}{1990}).

\bibitem[{\citenamefont{Pikovsky et~al.}(1997)\citenamefont{Pikovsky, Rosenblum,  and Kurths}}]{pikovsky01}
 \bibinfo{author}{\bibfnamefont{A.~S.} \bibnamefont{Pikovsky}}, 
\bibinfo{author}{\bibfnamefont{M.~G.} \bibnamefont{Rosenblum}},  \bibnamefont{and} 
  \bibinfo{author}{\bibfnamefont{J.}~\bibnamefont{Kurths}}, 
  \emph{\bibinfo{title}{Synchronization - A Universal Concept in Nonlinear
Sciences}} 
  (\bibinfo{publisher} {Cambridge University Press},
  \bibinfo{address}{Cambridge}, \bibinfo{year}{2001}).

\bibitem[{\citenamefont{Lakshmanan and Murali}(1996)}]{lakmurbook}
\bibinfo{author}{\bibfnamefont{M.}~\bibnamefont{Lakshmanan}},
  \bibnamefont{and} \bibinfo{author}{\bibfnamefont{K.}~\bibnamefont{Murali}},
  \emph{\bibinfo{title}{Chaos in Nonlinear Oscillators: Controlling and Synchronization}} 
  (\bibinfo{publisher} {World Scientific},
  \bibinfo{address}{Singapore}, \bibinfo{year}{1996}).

\bibitem[{\citenamefont{Senthilkumar et~al.}(2010)\citenamefont{Senthilkumar, Srinivasan, Murali, Lakshmanan and Kurths}}]{srini10}
\bibinfo{author}{\bibfnamefont{D.~V.} \bibnamefont{Senthilkumar}}, 
  \bibinfo{author}{\bibfnamefont{K.}~\bibnamefont{Srinivasan}},
  \bibinfo{author}{\bibfnamefont{K.}~\bibnamefont{Murali}}, 
  \bibinfo{author}{\bibfnamefont{M.}~\bibnamefont{Lakshmanan}}, \bibnamefont{and} 
  \bibinfo{author}{\bibfnamefont{J.}~\bibnamefont{Kurths}}, 
  \bibinfo{journal}{Phys. Rev. E} \textbf{\bibinfo{volume}{82}},
  \bibinfo{pages}{065201(R)} (\bibinfo{year}{2010}).

\bibitem[{\citenamefont{Srinivasan et~al.}(2011)\citenamefont{Srinivasan, Senthilkumar, Murali, Lakshmanan and Kurths}}]{srini11}
\bibinfo{author}{\bibfnamefont{K.}~\bibnamefont{Srinivasan}}, 
  \bibinfo{author}{\bibfnamefont{D.~V.} \bibnamefont{Senthilkumar}}, 
  \bibinfo{author}{\bibfnamefont{K.}~\bibnamefont{Murali}}, 
  \bibinfo{author}{\bibfnamefont{M.}~\bibnamefont{Lakshmanan}}, \bibnamefont{and} 
  \bibinfo{author}{\bibfnamefont{J.}~\bibnamefont{Kurths}}, 
  \bibinfo{journal}{ Chaos} \textbf{\bibinfo{volume}{21}},
  \bibinfo{pages}{023119} (\bibinfo{year}{2011}).

\bibitem[{\citenamefont{Rosenblum et~al.}(1997)\citenamefont{Rosenblum, Pikovsky, and Kurths}}]{rosenblum97}
\bibinfo{author}{\bibfnamefont{M.~G.} \bibnamefont{Rosenblum}}, 
  \bibinfo{author}{\bibfnamefont{A.~S.} \bibnamefont{Pikovsky}}, \bibnamefont{and} 
  \bibinfo{author}{\bibfnamefont{J.}~\bibnamefont{Kurths}}, 
  \bibinfo{journal}{ Phys. Rev. Lett.} \textbf{\bibinfo{volume}{78}},
  \bibinfo{pages}{4193} (\bibinfo{year}{1997}).

\bibitem[{\citenamefont{Voss}(2000)\citenamefont{Voss}}]{voss00}
\bibinfo{author}{\bibfnamefont{H.~U.} \bibnamefont{Voss}}, 
  \bibinfo{journal}{Phys. Rev. E} \textbf{\bibinfo{volume}{61}},
  \bibinfo{pages}{5115} (\bibinfo{year}{2000}).

\bibitem[{\citenamefont{Voss}(2001)\citenamefont{Voss}}]{voss01}
\bibinfo{author}{\bibfnamefont{H.~U.} \bibnamefont{Voss}}, 
  \bibinfo{journal}{Phys. Rev. Lett.} \textbf{\bibinfo{volume}{87}},
  \bibinfo{pages}{014102} (\bibinfo{year}{2001}).

\bibitem[{\citenamefont{Pyragas and Pyragiene}(2008)\citenamefont{Pyragas and Pyragiene}}]{pyragas08}
\bibinfo{author}{\bibfnamefont{K.}~\bibnamefont{Pyragas}}, \bibnamefont{and}
  \bibinfo{author}{\bibfnamefont{T.}~\bibnamefont{Pyragiene}}, 
  \bibinfo{journal}{Phys. Rev. E} \textbf{\bibinfo{volume}{78}},
  \bibinfo{pages}{046217} (\bibinfo{year}{2008}).

\bibitem[{\citenamefont{Masoller}(2001)\citenamefont{Masoller}}]{masoller01}
\bibinfo{author}{\bibfnamefont{C.}~\bibnamefont{Masoller}}, 
  \bibinfo{journal}{Phys. Rev. Lett.} \textbf{\bibinfo{volume}{86}},
  \bibinfo{pages}{2782} (\bibinfo{year}{2001}).

\bibitem[{\citenamefont{Wu and Zhu}(2003)\citenamefont{Wu and Zhu}}]{wu03}
\bibinfo{author}{\bibfnamefont{L.}~\bibnamefont{Wu}}, \bibnamefont{and}
  \bibinfo{author}{\bibfnamefont{S.}~\bibnamefont{Zhu}}, 
  \bibinfo{journal}{Phys. Lett. A} \textbf{\bibinfo{volume}{315}},
  \bibinfo{pages}{101} (\bibinfo{year}{2003}).

\bibitem[{\citenamefont{Li et~al.}(2004)\citenamefont{Li, Liao, and Wong}}]{li04}
\bibinfo{author}{\bibfnamefont{C.}~\bibnamefont{Li}}, 
  \bibinfo{author}{\bibfnamefont{X.}~\bibnamefont{Liao}}, \bibnamefont{and} 
  \bibinfo{author}{\bibfnamefont{K.}~\bibnamefont{Wong}}, 
  \bibinfo{journal}{ Physica D} \textbf{\bibinfo{volume}{194}},
  \bibinfo{pages}{187} (\bibinfo{year}{2004}).

\bibitem[{\citenamefont{Ciszak et~al.}(2004)\citenamefont{Ciszak, Marino, Toral and Balle}}]{ciszak04}
\bibinfo{author}{\bibfnamefont{M.}~\bibnamefont{Ciszak}}, 
  \bibinfo{author}{\bibfnamefont{F.}~\bibnamefont{Marino}}, 
  \bibinfo{author}{\bibfnamefont{R.}~\bibnamefont{Toral}}, \bibnamefont{and} 
  \bibinfo{author}{\bibfnamefont{S.}~\bibnamefont{Balle}}, 
  \bibinfo{journal}{ Phys. Rev. Lett.} \textbf{\bibinfo{volume}{93}},
  \bibinfo{pages}{114102} (\bibinfo{year}{2004}).


\bibitem[{\citenamefont{Pethel et~al.}(2003)\citenamefont{Pethel, Corron, Underwood and Myneni}}]{pethel03}
\bibinfo{author}{\bibfnamefont{S.~D.} \bibnamefont{Pethel}}, 
  \bibinfo{author}{\bibfnamefont{N.~J.} \bibnamefont{Corron}}, 
  \bibinfo{author}{\bibfnamefont{Q.~R.} \bibnamefont{Underwood}}, \bibnamefont{and} 
  \bibinfo{author}{\bibfnamefont{K.}~\bibnamefont{Myneni}}, 
  \bibinfo{journal}{ Phys. Rev. Lett.} \textbf{\bibinfo{volume}{90}},
  \bibinfo{pages}{254101} (\bibinfo{year}{2003}).

\bibitem[{\citenamefont{Blakely et~al.}(2008)\citenamefont{Blakely, Pruitt, and Corron}}]{blakely08}
  \bibinfo{author}{\bibfnamefont{J.~N.} \bibnamefont{Blakely}}, 
  \bibinfo{author}{\bibfnamefont{M.~W.} \bibnamefont{Pruitt}}, \bibnamefont{and} 
  \bibinfo{author}{\bibfnamefont{N.~J.} \bibnamefont{Corron}}, 
  \bibinfo{journal}{ Chaos} \textbf{\bibinfo{volume}{18}},
  \bibinfo{pages}{013117} (\bibinfo{year}{2008}).

\bibitem[{\citenamefont{Voss}(2002)\citenamefont{Voss}}]{voss02}
\bibinfo{author}{\bibfnamefont{H.~U.} \bibnamefont{Voss}}, 
  \bibinfo{journal}{Int. J. Bifurcation Chaos Appl. Sci. Eng.} \textbf{\bibinfo{volume}{12}},
  \bibinfo{pages}{1619} (\bibinfo{year}{2002}).

\bibitem[{\citenamefont{Taherion and Lai}(1999)\citenamefont{Taherion and Lai}}]{taherion99}
\bibinfo{author}{\bibfnamefont{S.}~\bibnamefont{Taherion}}, \bibnamefont{and}
  \bibinfo{author}{\bibfnamefont{Y.~C.} \bibnamefont{Lai}}, 
  \bibinfo{journal}{Phys. Rev. E} \textbf{\bibinfo{volume}{59}},
  \bibinfo{pages}{R6247} (\bibinfo{year}{1999}).

\bibitem[{\citenamefont{Lakshmanan and Senthilkumar}(2010)}]{lakdvsbook}
\bibinfo{author}{\bibfnamefont{M.}~\bibnamefont{Lakshmanan}},
  \bibnamefont{and} \bibinfo{author}{\bibfnamefont{D.~V.} \bibnamefont{Senthilkumar}},
  \emph{\bibinfo{title}{Dynamics of Nonlinear Time-Delay Systems}} 
  (\bibinfo{publisher} {Springer-Verlag},
  \bibinfo{address}{Berlin}, \bibinfo{year}{2010}).

\bibitem[{\citenamefont{Senthilkumar et~al.}(2005)\citenamefont{Senthilkumar and Lakshmanan}}]{dvs05}
\bibinfo{author}{\bibfnamefont{D.~V.} \bibnamefont{Senthilkumar}}, \bibnamefont{and} 
  \bibinfo{author}{\bibfnamefont{M.}~\bibnamefont{Lakshmanan}}, 
  \bibinfo{journal}{Phys. Rev. E} \textbf{\bibinfo{volume}{71}},
  \bibinfo{pages}{016211} (\bibinfo{year}{2005}).

\bibitem[{\citenamefont{Corron et~al.}(2005)\citenamefont{Corron, Blakely, and Pethel}}]{corron05}
\bibinfo{author}{\bibfnamefont{N.~J.} \bibnamefont{Corron}}, 
  \bibinfo{author}{\bibfnamefont{J.~N.} \bibnamefont{Blakely}}, \bibnamefont{and} 
  \bibinfo{author}{\bibfnamefont{S.~D.} \bibnamefont{Pethel}}, 
  \bibinfo{journal}{ Chaos} \textbf{\bibinfo{volume}{15}},
  \bibinfo{pages}{023110} (\bibinfo{year}{2005}).

\bibitem[{\citenamefont{Chua et~al.}(1987)}]{chuabook}
\bibinfo{author}{\bibfnamefont{L.~O.} \bibnamefont{Chua}},
  \bibinfo{author}{\bibfnamefont{C.~A.} \bibnamefont{Desoer}},
  \bibnamefont{and} \bibinfo{author}{\bibfnamefont{E.~S.}~\bibnamefont{Kuh}},
  \emph{\bibinfo{title}{Linear and Nonlinear
  Circuits}} (\bibinfo{publisher}{McGraw-Hill},
  \bibinfo{address}{Singapore}, \bibinfo{year}{1987}).

\bibitem[{\citenamefont{Ogorzatek}(1997)}]{ogo}
\bibinfo{author}{\bibfnamefont{M.~J.} \bibnamefont{Ogorzatek}},
  \emph{\bibinfo{title}{Chaos and Complexity in Nonlinear Electronic
  Circuits}} (\bibinfo{publisher}{World Scientific},
  \bibinfo{address}{Singapore}, \bibinfo{year}{1997}).

\bibitem[{\citenamefont{Chen and Ueta}(2002)}]{chenueta}
\bibinfo{author}{\bibfnamefont{G.}~\bibnamefont{Chen}},
  \bibnamefont{and} \bibinfo{author}{\bibfnamefont{T.}~\bibnamefont{Ueta}},
  \emph{\bibinfo{title}{Chaos in Circuits and System}} 
  (\bibinfo{publisher} {World Scientific},
  \bibinfo{address}{Singapore}, \bibinfo{year}{2002}).

\bibitem[{\citenamefont{Lakshmanan and Rajasekar}(2003)}]{mlsrbook}
\bibinfo{author}{\bibfnamefont{M.}~\bibnamefont{Lakshmanan}},
  \bibnamefont{and} \bibinfo{author}{\bibfnamefont{S.}~\bibnamefont{Rajasekar}},
  \emph{\bibinfo{title}{Nonlinear Dynamics: Integrability, Chaos and Pattern
  Formation}} 
  (\bibinfo{publisher} {Springer-Verlag},
  \bibinfo{address}{Berlin}, \bibinfo{year}{2003}).

\bibitem[{\citenamefont{Madan}(1993)}]{madan}
\bibinfo{author}{\bibfnamefont{R.~N.} \bibnamefont{Madan}},
  \emph{\bibinfo{title}{Chua's circuit: a paradigm for chaos}} 
  (\bibinfo{publisher}{World Scientific},
  \bibinfo{address}{Singapore}, \bibinfo{year}{1993}).

\bibitem[{\citenamefont{Chua et~al.}(1986)\citenamefont{Chua, Komuro, and Matsumoto}}]{chukom}
\bibinfo{author}{\bibfnamefont{L.~O.} \bibnamefont{Chua}}, 
  \bibinfo{author}{\bibfnamefont{M.}~\bibnamefont{Komuro}}, \bibnamefont{and} 
  \bibinfo{author}{\bibfnamefont{T.}~\bibnamefont{Matsumoto}}, 
  \bibinfo{journal}{ IEEE Trans. Circuits Syst.} \textbf{\bibinfo{volume}{CAS-33}},
  \bibinfo{pages}{1072} (\bibinfo{year}{1986}).

\bibitem[{\citenamefont{Kennedy}(1992)\citenamefont{Kennedy}}]{kenn1992}
\bibinfo{author}{\bibfnamefont{M.~P.} \bibnamefont{Kennedy}}, 
  \bibinfo{journal}{Frequenz} \textbf{\bibinfo{volume}{46}},
  \bibinfo{pages}{66} (\bibinfo{year}{1992}).

\bibitem[{\citenamefont{Pecora and Carroll}(1991)\citenamefont{Pecora and Carroll}}]{pecora91}
\bibinfo{author}{\bibfnamefont{L.~M} \bibnamefont{Pecora}}, \bibnamefont{and}
  \bibinfo{author}{\bibfnamefont{T.~L} \bibnamefont{Carroll}}, 
  \bibinfo{journal}{Phys. Rev. A.} \textbf{\bibinfo{volume}{44}},
  \bibinfo{pages}{2374} (\bibinfo{year}{1991}).

\bibitem[{\citenamefont{Endo and Chua}(1991)\citenamefont{Endo and Chua}}]{endo}
\bibinfo{author}{\bibfnamefont{T.}~\bibnamefont{Endo}}, \bibnamefont{and}
  \bibinfo{author}{\bibfnamefont{L.~O} \bibnamefont{Chua}}, 
  \bibinfo{journal}{Int. J. Bifurcation and Chaos} \textbf{\bibinfo{volume}{1}},
  \bibinfo{pages}{701} (\bibinfo{year}{1991}).

\bibitem[{\citenamefont{Vieira et~al.}(1991)\citenamefont{De Sousa Vieira, Lichtenberg, and Lieberman}}]{desousa}
\bibinfo{author}{\bibfnamefont{M.}~ \bibnamefont{De Sousa Vieira}}, 
  \bibinfo{author}{\bibfnamefont{A.~J.} \bibnamefont{Lichtenberg}}, \bibnamefont{and} 
  \bibinfo{author}{\bibfnamefont{M.~A.} \bibnamefont{Lieberman}}, 
  \bibinfo{journal}{Int. J. Bifurcation and Chaos} \textbf{\bibinfo{volume}{1}},
  \bibinfo{pages}{691} (\bibinfo{year}{1991}).

\bibitem[{\citenamefont{Hosokawa et~al.}(2001)\citenamefont{Hosokawa, Nishio, and Ushida}}]{hosokawa01}
\bibinfo{author}{\bibfnamefont{Y.}~\bibnamefont{Hosokawa}}, 
  \bibinfo{author}{\bibfnamefont{Y.}~\bibnamefont{Nishio}}, \bibnamefont{and} 
  \bibinfo{author}{\bibfnamefont{A.}~\bibnamefont{Ushida}}, 
  \bibinfo{journal}{ IEICE Trans. Fundamentals} \textbf{\bibinfo{volume}{E84-A}},
  \bibinfo{pages}{2288} (\bibinfo{year}{2001}).

\bibitem[{\citenamefont{Gotz et~al.}(1993)}]{gotz}
\bibinfo{author}{\bibfnamefont{M.}~\bibnamefont{Gotz}},
  \bibinfo{author}{\bibfnamefont{U.}~\bibnamefont{Feldmann}},
  \bibnamefont{and} \bibinfo{author}{\bibfnamefont{W.}~\bibnamefont{Schwarz}},
  \bibinfo{journal}{ IEEE Trans. Circuits Syst.-I} \textbf{\bibinfo{volume}{40}},
  \bibinfo{pages}{854} (\bibinfo{year}{1993}). 

\bibitem[{\citenamefont{Murali and Sinha}(2010)\citenamefont{Murali and Sinha}}]{murali07}
\bibinfo{author}{\bibfnamefont{K.} \bibnamefont{Murali}}, 
  \bibinfo{author}{\bibfnamefont{}~\bibnamefont{Sudeshna Sinha}},
  \bibinfo{journal}{Phys. Rev. E} \textbf{\bibinfo{volume}{75}},
  \bibinfo{pages}{025201(R)} (\bibinfo{year}{2007}).


%
%
%
%
%
%
%
%
%
%
%
\end{thebibliography}
\end{document}